# Milli-spinner thrombectomy

## Authors and Affiliations

Yilong Chang[1], Qi Li[1], Shuai Wu[1], Benjamin Pulli[2], Darren Samli[3], Paul Yock[4], Jeremy J. Heit[2], Ruike Renee Zhao[1]

[1]Department of Mechanical Engineering, Stanford University School of Engineering, Stanford, CA 94305, USA.

[2]Department of Radiology, Stanford University School of Medicine, Stanford, CA 94305, USA.

[3]Department of Pathology, Stanford University School of Medicine, Stanford, CA 94305, USA.

[4]Department of Bioengineering, Stanford University School of Engineering, Stanford, CA 94305, USA.

## Corresponding author

Correspondence to: Ruike Renee Zhao (rrzhao@stanford.edu)

## Abstract

Blockage of blood flow in arteries or veins by blood clots can lead to serious medical conditions[1, 2]. Mechanical thrombectomy (MT), minimally invasive endovascular procedures that utilize aspiration[3, 4], stent retriever[5, 6], or cutting mechanisms[7] for clot removal have emerged as an effective treatment modality for ischemic stroke, myocardial infarction, pulmonary embolism, and peripheral vascular disease[8-10]. However, state-of-the-art MT technologies still fail to remove clots in approximately 10% to 30% of patients[11-14], especially when treating large-size clots with high fibrin content [15, 16]. In addition, the working mechanism of most current MT techniques results in rupturing or cutting of clots which could lead to clot fragmentation[17-19] and distal emboli[20, 21]. Here, we report a new MT technology based on an unprecedented mechanism, in which a milli-spinner mechanically debulks the clot by densifying its fibrin fiber network and discharging red blood cells (RBCs) to significantly reduce the clot volume and facilitate complete clot removal. This mechanism is achieved by the spin-induced compression and shearing of the clot. With the computational fluid dynamics guided structural design of the milli-spinner, we demonstrate its effective clot-debulking performance with clot volumetric reduction of up to 90% on various sizes of clots and on diverse clot compositions ranging from RBC-rich soft clots to fibrin-rich tough clots. Milli-spinner MT in both *in-vitro* pulmonary and cerebral artery flow models and *in-vivo* swine models demonstrate high-fidelity revascularization. The milli-spinner MT is the first reported mechanism that directly modifies the clot microstructure to facilitate clot removal, which also results in markedly improved MT efficacy compared to the existing MT mechanisms that are based on clot rupturing and cutting. This technology introduces a unique mechanical way of debulking and removing clots for future MT device development, especially for treatment of ischemic stroke, pulmonary emboli, and peripheral thrombosis.

**Introduction**

The blockage of blood flow in arteries or veins by clots can lead to critical medical conditions such as acute ischemic stroke, myocardial infarction, pulmonary embolism, peripheral vascular disease, and deep vein thrombosis (**Fig. 1a**). Mechanical thrombectomy (MT), a minimally invasive endovascular technology, has emerged as an effective clinical strategy to treat vascular occlusion. Most state-of-the-art MT techniques utilize aspiration and stent retrievers (**Fig. 1b**) to engage and extract clots, which results in vascular recanalization when successful and an improvement in disease[22-24]. Other MT techniques utilize a cutting mechanism to break clots into fragments for blood flow restoration in occluded vessels, but these approaches are generally restricted to clots caused by atherosclerotic disease only[25, 26]. Clinically, MT is an effective treatment for vascular occlusion[27-33]. However, current MT techniques still encounter failures especially when treating large and/or fibrin rich clots. Additionally, these techniques can induce rupturing and cutting of clots that cause clot fragmentation, leading to distal vessel occlusion[34] and unfavorable clinical outcomes[35, 36]. This failure is particularly true for acute ischemic stroke, in which the distal emboli can cause an increase in disability and mortality due to failed complete recanalization[37]. Considering these challenges, we develop a new MT technology based on an unprecedented mechanism, which involves a milli-spinner that mechanically debulks the clot by densifying its fibrin fiber network and discharges red blood cells (RBCs) to significantly reduce the clot volume for complete clot removal.

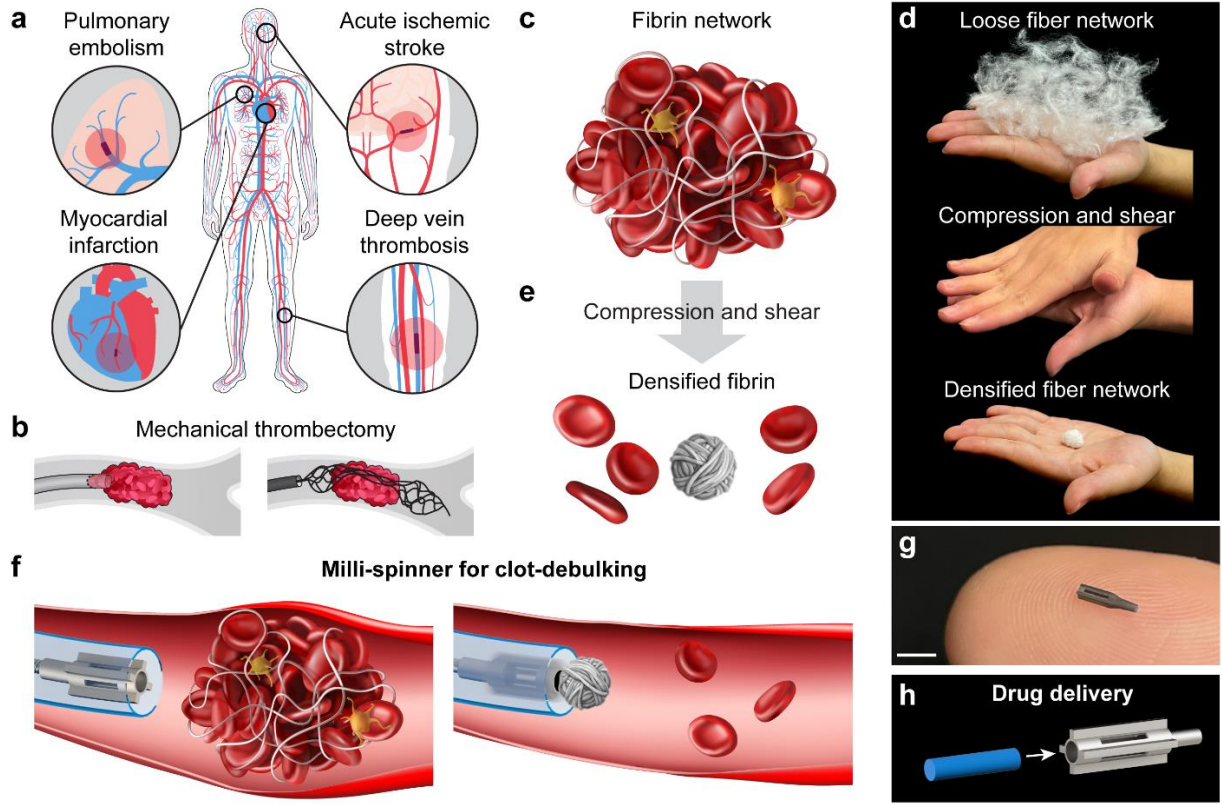

**Fig. 1 A multifunctional milli-spinner for effective mechanical thrombectomy (MT) to treat various medical conditions**. **a**, Example medical conditions due to blood vessel occlusion by clots. **b**, Schematics of commonly used MT techniques (left: aspiration-based, right: stent retriever-based). **c**, Schematic of a blood clot mainly composed of fibrin network and red blood cells (RBCs). **d**, Concept of densifying loose fiber network by coupled compression and shear loadings for drastic volumetric reduction. **e**, Clot size reduction by the densification of the fibrin network and discharging RBCs under compression and shear loadings. **f**, Schematic displaying milli-spinner debulking a clot. **g**, Photo of a printed milli-spinner with an outer diameter (OD) of 1.2 mm. Scale bar: 3 mm. **h**, Schematics showing multifunctional milli-spinner for drug delivery.

As illustrated in **Fig. 1c**, a clot is primarily composed of RBCs trapped within a fibrin network[38]. The milli-spinner debulks the clot through densifying the fibrin network that drastically reduces the clot volume, like rubbing a cotton ball. As shown in **Fig. 1d,** when the loose cotton fiber network is pressed and rubbed between two palms, the compression and shear forces densify and entangle the cotton fiber network, which results in a significant volume reduction. The same concept is adopted to densify the fibrin network for clot-debulking (**Fig. 1e**), where the coupled compression and shear forces are provided by the spinning motion of a rationally designed endovascular milli-spinner (**Fig. 1f**). During the clot-debulking, the milli-spinner generates a highly localized suction by spinning, which firmly presses the clot against the milli-spinner's front surface to enhance the shear force on the clot for fibrin densification, in the meantime, squeezing out the RBC from the fibrin network. This process leaves behind a highly densified fibrin core with clot volume reduction of up to 90%. The milli-spinner MT is the first reported mechanism that directly modifies the clot microstructure to facilitate clot removal, which also results in a much-improved outcome compared to the existing MT mechanisms that are based on clot rupturing and cutting. **Fig. 1g** shows the size of a printed milli-spinner on a fingertip. The spinner can be manufactured in various sizes for different blood vessels. Additionally, the lumen of the milli-spinner can be exploited for drug storage and targeted delivery (**Fig. 1h**).

In the following study, we first quantitatively investigate the geometric features of the milli-spinner for optimized clot-debulking performance through computational fluid dynamics (CFD) simulations. With the optimized design, the milli-spinner's clot-debulking efficacy is systematically studied for clots of various sizes and compositions ranging from RBC-rich soft clots to fibrin-rich tough clots. Fluoroscopy-guided studies are then conducted in the pulmonary artery and cerebral artery flow models, which demonstrate the milli-spinner's ultrafast clot-debulking and removal. Finally, *in-vivo* thrombectomy experiments are carried out in the renal and facial arteries of swine models to further evaluate the milli-spinner's performance in removing clots.

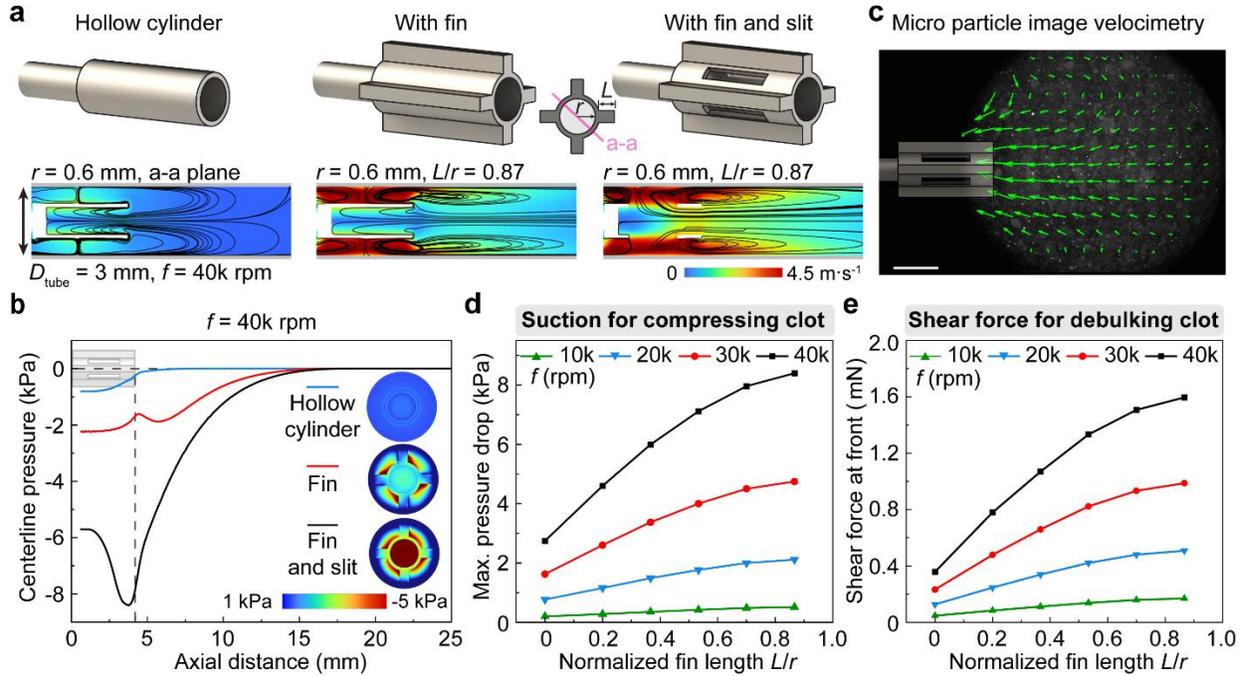

**Fig. 2 Milli-spinner geometric optimization for suction and shear through computational fluid dynamics simulation. a**, Different milli-spinner designs: hollow cylinder, hollow cylinder with fin design, hollow cylinder with fin and slit design. The milli-spinner designs have hollow cylinder inner diameter as $r$ and the two designs with fin have fin length as $L$. Corresponding velocity contours with streamlines of three milli-spinner designs on the a-a plane ($r = 0.6$ mm, normalized fin length $L/r = 0.87$) illustrate flow towards the milli-spinner's cavity at a spinning frequency $f = 40$k rpm in a tube with 3 mm diameter ($D_{tube}$ = 3 mm). **b**, Centerline pressure distribution with respect to the axial distance of the three milli-spinner designs in (a) at $f = 40$k rpm. Pressure contour at the front surface plane of the milli-spinner designs in (a) at $f = 40$k rpm. **c**, Micro particle image velocimetry of the fin and slit design at $f = 1.6$k rpm in a 36 mm × 10 mm × 10 mm tank, illustrating the design's suction capability. Scale bar: 2 mm. **d**, Maximum centerline pressure drop of the with fin and slit design with respect to $L/r$ at varied $f$. **e**, Shear force at the front surface of the with fin and slit design with respect to $L/r$ at varied $f$.

**Structural optimization of the milli-spinner for enhanced compression and shear**

The clot-debulking mechanism (Fig. 1f) of the milli-spinner is attributed to the shear force it generates on the clot. In an endovascular environment, a continuous shear force on a clot is most effectively produced when the clot is in contact with a rotating component. The shear force can be further enhanced through increasing the friction by pressing the clot onto the rotation component. To design the milli-spinner's geometric feature that maximizes the compression and shear force on the clot for optimal clot-debulking, CFD is used to investigate the hydrodynamics of different spinning milli-spinner designs (See Method section for more CFD simulation details).

Starting from a hollow cylinder milli-spinner, the most intuitive design to generate suction for clot compression, we study the flow velocity field, pressure distribution, and surface shear to evaluate the clot-debulking capability of different milli-spinner designs. As shown in **Fig. 2a**, a hollow cylinder (inner diameter $r = 0.6$ mm) spinning inside a tube (tube diameter $D_{tube}$ of 3 mm) can generate a localized suction, demonstrated by the fluid flow velocity vector towards the milli-spinner front hole and the pressure drop in the lumen along the centerline (**Fig. 2b**). Adding fin features (the second design in **Fig. 2a**) is shown to enhance the suction by a larger negative pressure within the lumen and in front of the milli-spinner. The design with four evenly spaced straight fins extending radially from the hollow cylinder is used here. The suction can be drastically enhanced by adding slits on the cylinder body between fins (the third design in **Fig. 2a**). This is evidenced by both the high-velocity gradient and the significantly increased pressure drop in front and in the lumen of the milli-spinner. The peak centerline pressure drop is nearly eight times that of the hollow cylinder design. This extraordinary suction, which is directly correlated to the clot compression capability, is due to the hole-slit combination that allows for flow circulation by sprinting flow into the milli-spinner through the front hole and spinning out from the slit. **Fig. 2c** shows the micro particle image velocimetry (Micro-PIV) result of the milli-spinner design with fin and slit spinning at 1.6k rpm, illustrating flow towards the milli-spinner front hole. More experimental details are provided in the Method section. The Micro-PIV results validate the CFD simulation.

For the novel milli-spinner design with the combination of hole, fin, and slit, the CFD simulation in **Fig. 2d** and **Fig. 2e** show that both the maximum suction pressure and the shear force at the front plane of the milli-spinner increase monotonically with the normalized fin length $L/r$ and the spinning frequency $f$. To ensure a good suction and shear for effective clot-debulking, a milli-spinner with $L/r = 0.87$ is adopted for the following studies and is referred to as the milli-spinner for convenience. The size of the milli-spinner can also be scaled to fit the size of the target blood vessel.

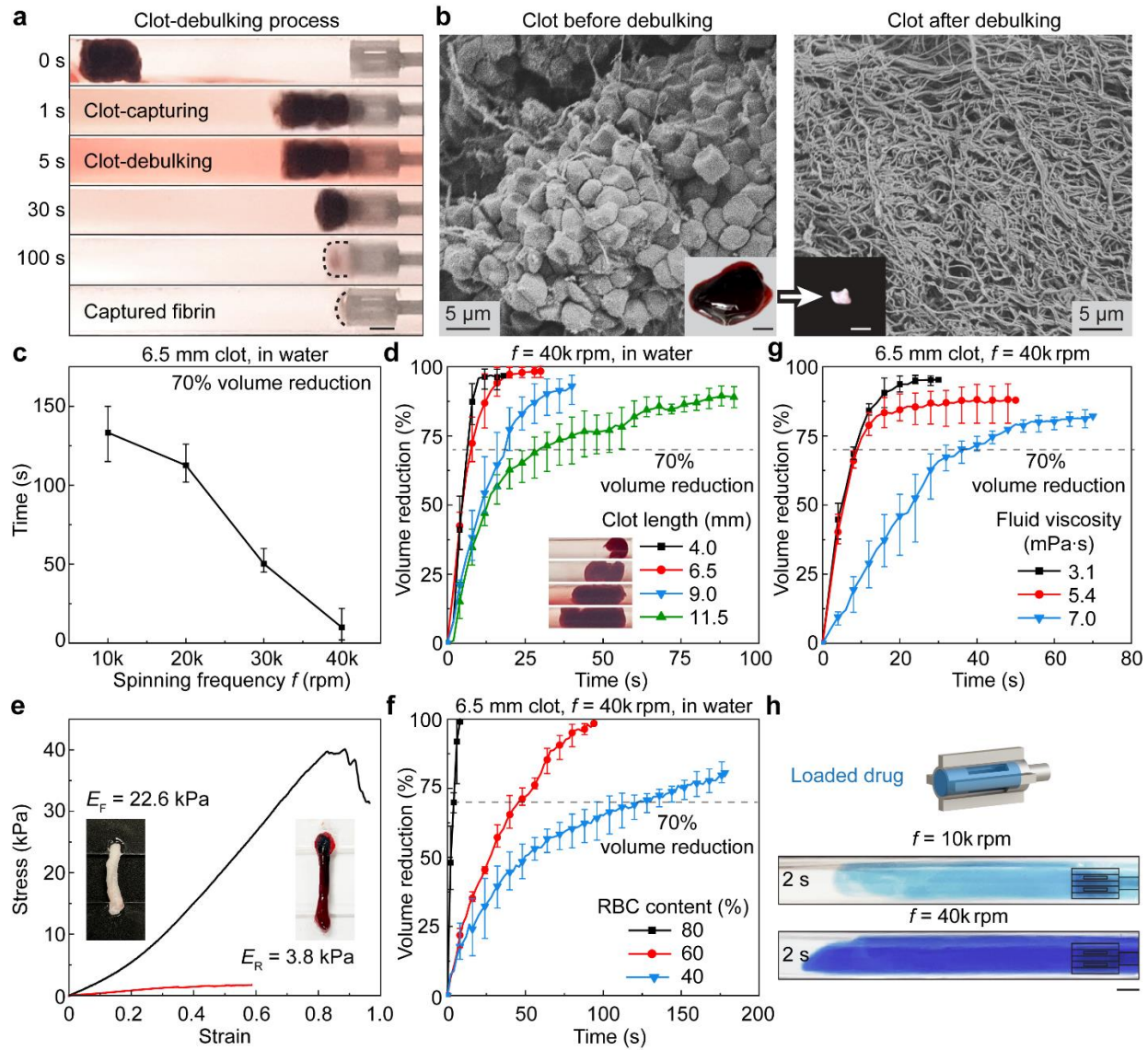

**Fig. 3 Milli-spinner clot-debulking mechanism and efficiency characterization on clots with various sizes, RBC contents, and in fluid with different viscosities. a**, Experimental demonstration of the milli-spinner clot-debulking a 6.5 mm long whole blood formed clot at $f$ = 10k rpm. The milli-spinner has an OD of 2.5 mm, and the tube has a diameter of 3 mm. The following characterization studies are conducted inside the same tube with the same milli-spinner. **b**, Comparison of SEM images and pictures of the whole blood formed clot before and after debulking. **c**, Debulking time required to reach 70% volume reduction of 6.5 mm long whole blood formed clots at $f$ = 10k, 20k, 30k, and 40k rpm. **d**, Milli-spinner clot-debulking efficiency on 4, 6.5, 9, and 11.5 mm long whole blood formed clots at $f$ = 40k rpm. **e**, Stiffness characterization of white fibrin clot and whole blood formed clot through the tensile test. **f**, Milli-spinner clot-debulking efficiency on 6.5 mm long whole blood formed clots at $f$ = 40k rpm in fluid with viscosities of 3.1, 5.4, and 7.0 mPa·s. **g**, Milli-spinner clot-debulking efficiency on clots with RBC contents of 80%, 60%, and 40% at $f$ = 40k rpm. **h**, Spinning-controlled targeted drug release of milli-spinner. Scale bars: 2 mm unless specified.

**Clot-debulking mechanism and efficacy characterization**

**Fig. 3a** shows the experimental clot-debulking by the milli-spinner (Mechanism illustrated in Fig. 1f), where the clot volume reduces to only 10% of its initial volume in less than 100 s, which is accompanied by a clot color change from dark red to white. In **Fig. 3b**, the scanning electron microscopy images (SEM) (See Method section for clot SEM imaging procedure) of the clot before and after debulking demonstrate that the milli-spinner changes the composition of the clot through fibrin densification. The RBCs within the clot are discharged from the clot whereas the densified fibrin is captured by the milli-spinner. This mechanism is the first one reported to remove clots by manipulating clot composition. The clot-debulking speed increases significantly with the spinning frequency, as shown in **Fig. 3c**. A batch of 6.5 mm long whole blood formed clot is debulked under different spinning frequencies, from 10k to 40k rpm. At 40k rpm, the clot shows a 70% volume reduction in 10 s. This ultra-fast clot-debulking is due to the drastically increased suction and shear force exerted on the clot when the spinning frequency is high. The clot-debulking efficiency is characterized with respect to clot sizes, shown in **Fig. 3d**. In general, under 40k rpm, 4 mm to 11.5 mm whole blood formed clots can be debulked to less than 30% of its volume in roughly 5 to 40 s and at least 85% volume reduction is achieved for all tested clots. Note that the typical clot length for acute ischemic stroke ranges from 4 mm to 21.5 mm[39].

Clot composition is highly variable and is an important factor that affects the efficacy of the existing MT. Tough clots such as arterial clots formed under high blood flow or chronic clots with high fibrin content are extremely difficult to address by the clot rupture and cutting-based mechanisms of existing MT techniques. As shown by the clot tensile tests in **Fig. 3e**, the stiffness and toughness of a clot vary significantly based on the fibrin content. A fibrin rich clot (or white clot) can be 10 times stiffer and orders of magnitude tougher than an RBC-rich clot[40]. Our measurements show a fibrin clot has a Young's modulus of 22.6 kPa, and a toughness of 19.3 kJm$^{-3}$, while an RBC-rich clot only has a Young's modulus of 3.8 kPa and a toughness of 0.6 kJm$^{-3}$ (See Method section for white clot fabrication process and detailed information on clot mechanical property characterization through the tensile test). To demonstrate the milli-spinner's superior clot-debulking capability for stiff and tough clots, we conducted debulking tests on 6.5 mm long clots with 80%, 60%, and 40% RBC contents (**Fig. 3f**). Besides the ultrafast debulking of high RBC content (80%) clot in 3 s, the 60% and 40% RBC contents clots can reduce more than 70% of their volume in around 50 and 130 s, respectively, under a 40k rpm spinning frequency. This remarkable debulking performance of the milli-spinner provides a new strategy to treat tough clots in a fast and reliable way, which the current MT devices fail to achieve.

The effect of fluid viscosity on the milli-spinner's performance is also evaluated (blood has viscosity ranges from 3.5 to 5.5 mPa·s[41]). As shown in **Fig. 3g**, in fluid with a viscosity of 3.1 mPa·s, a

6.5 mm long whole blood formed clot (RBC content around 95%) can be reduced to 70% of its initial volume in less than 10 s under 40k rpm spinning frequency and ultimately reach a volume reduction of more than 90%. When debulking in fluid with a viscosity of 5.4 and 7.0 mPa·s, 70% volume reduction of the clots can be achieved within 10 and 40 s, respectively, and both eventually reach a volume reduction of more than 80%. The viscous fluid preparation and characterization details are provided in the Method section.

Last, the milli-spinner lumen can also be used to store and deliver drugs such as tissue plasma activators (tPA) to clot to potentially facilitate even faster clot removal (**Fig. 3h**). To demonstrate this concept, dye is stored in milli-spinner cavity representing drug. The drug release speed can be well controlled by the spinning frequency. For a fast spinning of 40k rpm, an intense release of the drug, represented by the dark blue color is shown in contrast to the gradual release of blue dye at 10k rpm spinning frequency represented by the light blue color. See the Method section for the milli-spinner with drug preparation process.

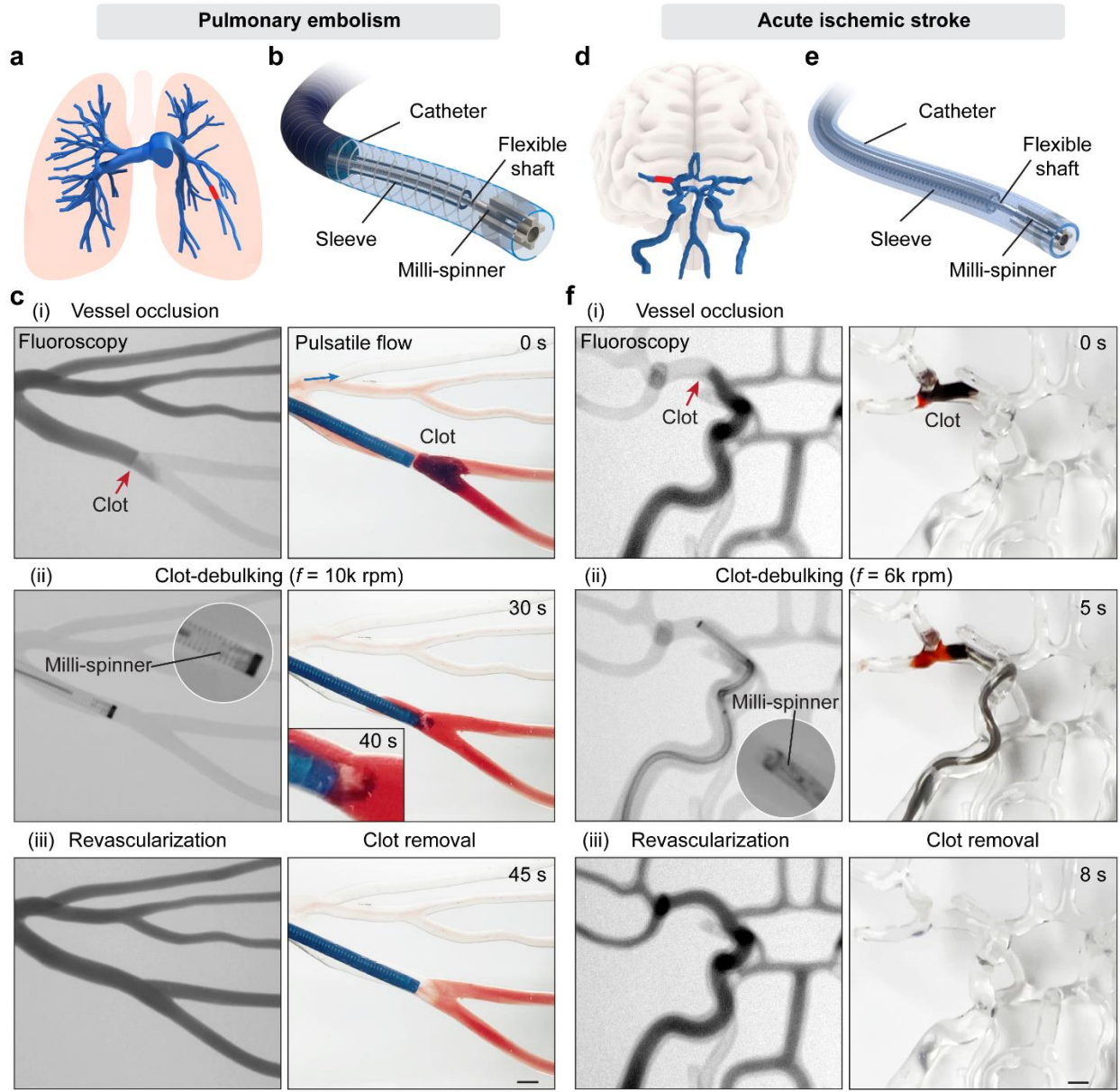

**Fig. 4 Fluoroscopy guided *in-vitro* clot-debulking and clot removal in the pulmonary artery and cerebral artery flow models. a**, Schematic of pulmonary embolism, clot (red) blocking pulmonary artery branch. **b**, Schematics of milli-spinner system setup utilized in the clot-debulking experiment in pulmonary artery flow model. **c**, Fluoroscopic images and photos show milli-spinner *in-vitro* clot-debulking in pulmonary artery flow model with 2.5 mm OD milli-spinner at $f$ = 10k rpm. Fluoroscopic images and photos show (i) occlusion by whole blood formed clot at vessel bifurcation point; (ii) significant clot volume reduction with clot revealing white fibrin after 30 s of clot-debulking. Milli-spinner is completely positioned inside the catheter; (iii) revascularization of the pulmonary artery flow model after 45 s of clot-debulking. **d**, Schematics of acute ischemic stroke, clot (red) blocking the middle cerebral artery section of cerebral artery. **e**, Schematic of milli-spinner setup utilized in the clot-debulking experiment in the cerebral artery flow model. **f**, Fluoroscopic images and photos show milli-spinner *in-vitro* clot-debulking in cerebral artery flow model with 1.2 mm OD milli-spinner at $f$ = 6k rpm. Fluoroscopic images and photos show (i) occlusion by whole blood formed clot at middle cerebral artery section of cerebral artery flow model; (ii) significant

clot volume reduction after 5 s of clot-debulking. (iii) revascularization of the cerebral artery flow model after 8 s of clot-debulking. Scale bars: 5 mm.

**Fluoroscopy-guided clot-debulking in the pulmonary artery and cerebral artery flow models**

*In-vitro* clot-debulking tests under fluoroscopic guidance are conducted in the pulmonary artery and cerebral artery flow models to evaluate the effectiveness of milli-spinner, potentially mapping to its performance in treating pulmonary embolism and acute ischemic stroke. Here, the 3D printed milli-spinners are radio-opaque for visibility under fluoroscopic guidance (See Method section for radio-opaque milli-spinner material property characterization).

Pulmonary embolism is a blockage of blood flow by clots in the pulmonary artery (**Fig. 4a**). The milli-spinner system (**Fig. 4b**) is tested in a pulmonary artery flow. A whole blood formed clot (with a length of 13 mm) travels with the flow (blue arrow in **Fig. 4c-i**) in the model and stops at a bifurcation point (7.8 mm in diameter), which causes a vessel occlusion that mimics pulmonary embolism (**Fig. 4c-i**). The angiogram shows the arrest of contrast flow, which indicates the clot location. It is important to note that the operator of the milli-spinner system has access to clot location solely through fluoroscopic guidance to better mimic actual MT procedures. A catheter is navigated under fluoroscopic guidance and delivered to the proximal end of the clot. The milli-spinner (OD = 2.5 mm) is then inserted and stopped at the distal end of the catheter to ensure its entire body is kept inside the catheter to prevent potential vessel damage (**Fig. 4c-ii**). During the debulking process, manual aspiration through the catheter is applied to strongly press the clot against the milli-spinner front surface for a much-enhanced shear force to improve clot-debulking efficiency. After 1 minute, the milli-spinner and catheter are removed from the flow model, followed by an injection of the contrast agent to illustrate revascularization (**Fig. 4c-iii**). The recorded process shows that the clot volume is reduced by more than 50%, accompanied by the appearance of white fibrin, after just 30 s of clot-debulking (**Fig. 4c-ii**). After 45 s, the clot is reduced to an even smaller size that can be directly aspirated into the catheter for clot removal (**Fig. 4c-iii**). In contrast to state-of-the-art MT technologies requiring cutting or rupturing of the clot which increases the risk of failure and clot fragmentation, the milli-spinner can debulk the clot to a smaller size, for an easier, safer, and more reliable treatment.

Acute ischemic stroke (AIS) (**Fig. 4d**) is a life-threatening medical emergency caused by a clot blocking the blood flow in the cerebral artery. It has been reported by studies and clinical trials that the single use of a device achieving complete revascularization (first-pass effect) can maximize the likelihood of good outcomes[42, 43]. However, current MT technologies often require multiple MT attempts to remove the offending clot completely,[44, 45] and the first-pass effect only achieved in 25~40% of patients who undergo mechanical thrombectomy[46-48]. Here, we demonstrate milli-spinner clot-debulking and clot removal in the middle cerebral artery section in a single use of device to showcase the milli-spinner's potential for treating AIS for a much improved first-pass effect. A milli-spinner (OD = 1.2 mm) is adopted to accommodate a 6 French aspiration catheter that is commonly used for treating AIS (**Fig. 4e**) is manually loaded into the cerebral artery and pushed by flow to the middle cerebral artery region (~3 mm in diameter) causing vessel occlusion. The occlusion location is identified by contrast agent injection under fluoroscopic guidance (**Fig. 4f-i**). The milli-spinner is then delivered to the clot site with its whole body being kept inside the catheter. Aspiration through a vacuum pump (Gomco Pump, Allied Healthcare Products, Inc.) is applied to press the clot against the milli-spinner front surface to escalate the shearing of the clot for fast debulking. After merely 5 s of clot-debulking, the clot volume is significantly reduced, as shown in **Fig. 4f-ii**. The clot is completely removed after 8 s and contrast is then injected through the catheter showing revascularization with a single use of milli-spinner device (**Fig. 4f-iii**). A remarkable 100% success rate is achieved in over 500 *in-vitro* clot-debulking tests in both flow models.

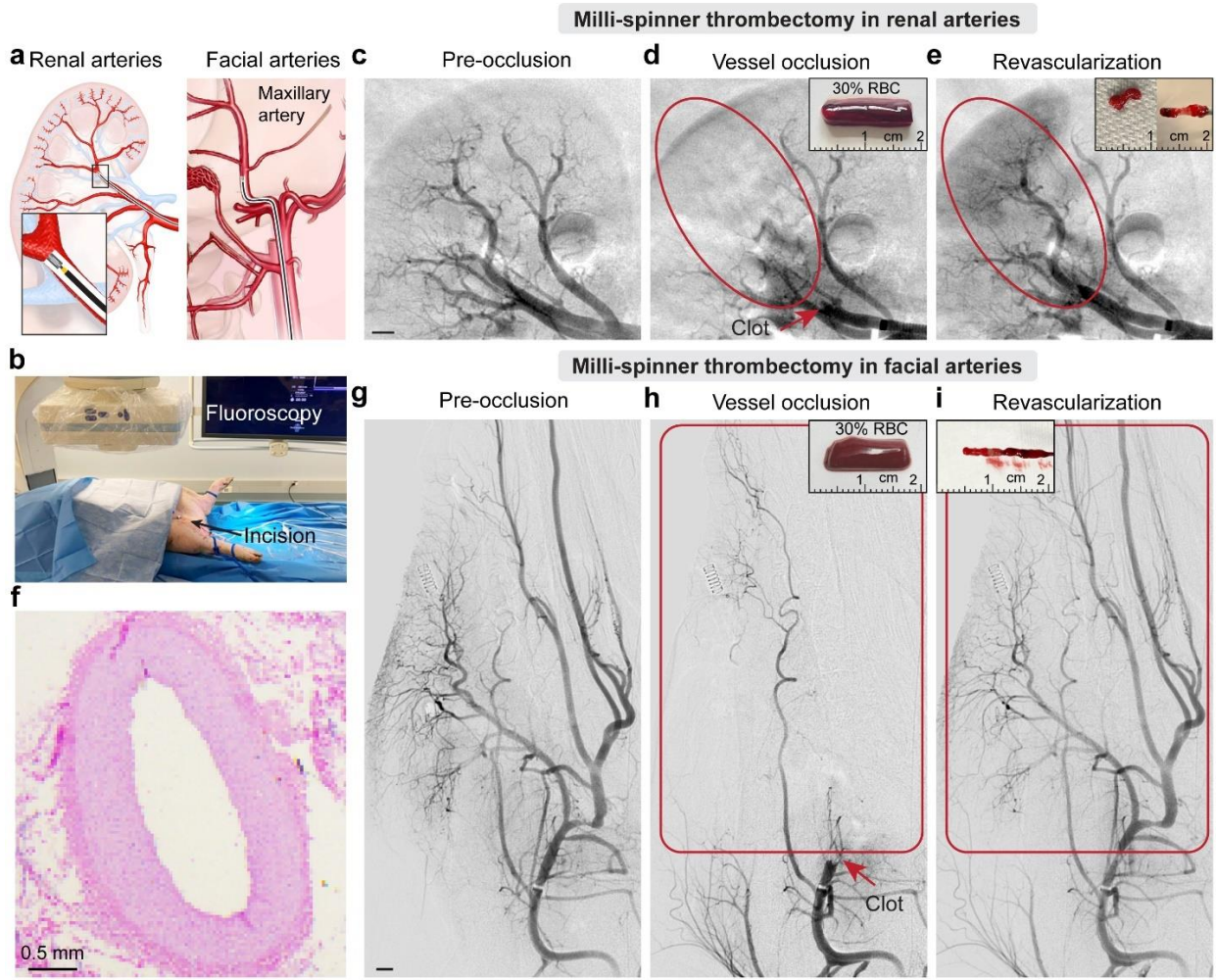

**Fig. 5 Fluoroscopic guided *in-vivo* milli-spinner thrombectomy in swine renal arteries and facial arteries. a**, Schematics of milli-spinner system inside the swine renal and facial arteries. **b**, Swine model and fluoroscopy setup of the *in-vivo* milli-spinner thrombectomy. **c**, Digitally subtracted angiography (DSA) of swine renal arteries before inducing vessel occlusion. **d**, DSA shows occlusion of renal arteries after injection of a clot (30% RBC content clot with a diameter of 5 mm and length of 15 mm). **e**, DSA shows revascularization of the renal arteries after 2 minutes of clot-debulking (milli-spinner OD = 1.5 mm, $f$ = 10k rpm). Milli-spinner after operation showing densified residual fibrin clot. **f**, Histology of vessel slice at milli-spinner tip position depicting intact endothelium layer. **g**, DSA of swine facial arteries before inducing vessel occlusion. **h**, DSA shows occlusion of the facial arteries after injection of a clot (30% RBC content clot with a diameter of 5 mm and length of 15 mm). **i**, DSA shows revascularization of the facial arteries after 2 minutes of clot-debulking ($f$ = 10k rpm, milli-spinner OD = 1.3 mm). Milli-spinner after operation showing densified residual fibrin clot. Scale bars: 5 mm unless specified.

**Milli-spinner thrombectomy study in swine models**

Milli-spinner thrombectomy is conducted in swine models to test clot removal performance and safety of milli-spinner when operating in an endovascular environment (**Fig. 5a**) under fluoroscopic guidance (**Fig. 5b**). More specifically, the experiments are designed for testing milli-spinner's clot removal capability in an environment that best mimics acute ischemic stroke condition. Thus, swine renal arteries are chosen as the testing locations considering the vessel size similarity with human middle cerebral arteries. The facial arteries of swine are selected as another test site to more accurately replicate the operational conditions within the tortuous human cerebral arteries.

The digitally subtracted angiography (DSA) of the swine renal arteries before vessel occlusion is captured (**Fig. 5c**). The vessel's size is measured, which shows a lumen diameter of around 3 mm (the human middle cerebral arteries range from 2 to 5 mm in diameter[49]). A 30% RBC content clot is syringe-injected through a long sheath into the renal arteries and results in vessel occlusion. The occlusion location is identified by injecting contrast agent into the renal vascular system (**Fig. 5d**). An aspiration catheter is navigated to the occlusion site followed by milli-spinner system delivery through the catheter. The front surface of the milli-spinner is aligned with the catheter opening for safety and optimal clot-debulking functionality. After 2 minutes of clot-debulking aided with vacuum aspiration, the milli-spinner and aspiration catheter are retracted out of the swine model. The milli-spinner is observed to have fibrin entangled on itself (**Fig. 5e**). Contrast agent is injected into the treated vascular system and flows through the previously occluded vessel and downstream area demonstrating complete revascularization of the vessel through a single use of milli-spinner device.

To compare with a state-of-the-art MT device, a direct aspiration thrombectomy targeting occlusion at the same vessel location is conducted. The direct aspiration technique fails to remove the clot after a single operation run. To visualize the clot-debulking process *in-vivo*, radio-opaque clot is utilized (See Method section for radio-opaque clot fabrication process). It is shown that the radio-opaque clot is gradually reducing in size and eventually is removed into the catheter realizing clot removed in about 13 s.

Histology analysis is conducted in milli-spinner treated renal arteries to assess the safety of the milli-spinner. This specific studied location is coincident with the catheter tip location, which is the milli-spinner tip position during the debulking process. The histological image of the vessel slice reveals an intact endothelium layer and reveals no signs of vessel damage (**Fig. 5f**) (See Method section for histology information).

Considering the tortuosity of the human cerebral arteries, it is important to evaluate the milli-spinner thrombectomy functionality in a tortuous vessel environment. Thus, the swine facial artery is chosen as a testing site since it possesses tortuosity. DSA of the facial arteries before vessel occlusion is captured (**Fig. 5g**). The occlusion in facial arteries induced by injection of a 30% RBC content clot is spotted by contrast agent flow (**Fig. 5h**). Milli-spinner is delivered to the occlusion location through an aspiration catheter and spins for 2 minutes with vacuum aspiration. Followed by retraction of the milli-spinner and aspiration catheter out of the swine model, the milli-spinner is observed to have fibrin entangled on itself (**Fig. 5i**). Contrast agent is injected into the treated facial arteries which demonstrates restored flow through the occlusion location and its downstream area. An additional *in-vivo* milli-spinner thrombectomy test conducted in the swine facial arteries demonstrates successful clot removal by milli-spinner. A summary of the milli-spinner thrombectomy success rate evaluated based on thrombolysis in cerebral infarction[50] (TICI) score shows the milli-spinner is capable of achieving complete revascularization (TICI 3) through a single use of milli-spinner in 80.6% of the time in treating tough clots. This number is expected to be higher when treating high RBC soft clots.

**Conclusion**

In this work, we report an endovascular milli-spinner that, under spinning, is capable of debulking clots through blood clot compression and shearing that results in significant clot volume reduction and successful clot removal. The clot-debulking process is attributed to the densification of the fibrin microstructure and discharging of RBCs. The clot-debulking efficacy of milli-spinner is demonstrated and quantitatively investigated on clots with various sizes and RBC contents, showing effective volume reductions for both large and tough clots. *In-vitro* milli-spinner thrombectomy tests in flow models under fluoroscopic guidance are conducted, showing the milli-spinner's capability in treating vessel occlusion in flow models. The *in-vivo* milli-spinner thrombectomy tests in both renal and facial arteries of swine models show revascularization through a single use of a milli-spinner device. The milli-spinner MT is the first reported mechanism that directly modifies the clot microstructure to facilitate clot removal, which also results in a much-improved outcome compared to the existing MT mechanisms that are based on clot rupturing and cutting. We envision that this novel milli-spinner mechanical thrombectomy can be the next generation of technology to significantly increase the success rate of treating acute ischemic stroke, pulmonary embolism, myocardial infarction, and peripheral vascular diseases.

**Method**

**Milli-spinner fabrication**

*Milli-spinner with OD = 2.5 mm*
Milli-spinner was printed using a commercialized stereolithography printer Form 3+ (Formlabs Inc., USA) with Formlabs Grey resin with a printing resolution of 25 µm.

*Radio-opaque milli-spinner (OD = 1.5, 1.3, 1.2 mm)*
Radio-opaque milli-spinners were fabricated using a customized digital light processing printer for their small feature size. A 385 nm UV-LED light projector (PRO4500, Wintech Digital Systems Technology Corp., USA) combined with an optical lens to enable a printing resolution of 14 µm. The printing resin was prepared from 69.5wt% Formlabs BioMed Durable resin, 30wt% barium sulfate (1-4 µm average, Thermo Fisher Scientific, USA), and 0.5wt% iron oxide (300 nm, Alpha Chemicals., USA). The resin was mixed by a planetary mixer (AR-100, Thinky, USA) at 2000 rpm for 30 s to ensure homogenization before printing. The printing light intensity was 2.72 mW·cm$^{-2}$ with a curing time of 2.5 s for each layer of 20 µm.

**Blood clot preparation**

*Whole blood formed clot*
Arterial blood was obtained from the femoral artery of pig species at a licensed facility (Stanford University School of Medicine). Spontaneous coagulation was initiated by collecting whole blood in a 50 ml centrifugation tube, left stationary for approximately 5 hours. The formed whole blood clot was stored in a fridge at 5°C. Prior to testing, the clot in the 50 ml centrifugation tube was cut into smaller clot pieces.

*Fibrin white clot*
Freshly extracted arterial blood was centrifugated at 1100 g for 15 minutes for the separation of blood. Centrifuged blood was left at room temperature for 5 hours for spontaneous coagulation. White fibrin clot was formed from plasma coagulation and was stored in a fridge at 5 °C. Prior to testing, the white fibrin clot was cut into smaller pieces for testing.

*Clots with varied RBC contents*
Porcine whole blood anticoagulated with 3.8wt% sodium citrate at a 9:1 volume ratio was purchased from Animal Technologies, Inc. The anticoagulated whole blood was centrifuged at 1100 g in a 50 ml centrifugation tube for 15 minutes for the separation of blood constitutes. Separated RBCs and citrated plasma were extracted separately and mixed at volume ratios of 8:2, 6:4, 4:6, and 3:7 for the making of clots with 80%, 60%, 40%, and 30% RBC contents. The mixtures were subsequently coagulated with the addition of 2.06wt% calcium chloride at the volume ratio of 9:1. The clots were water-bathed at 37 °C for approximately 1 hour and were stored in the fridge at 5°C overnight.

*Radio-opaque clot*

Clots were prepared following the clots with varied RBC contents method above. Before the addition of calcium chloride for coagulation, barium sulfate was added to RBC and plasma mixture at a concentration of 1.25 g per 30 ml of the mixture. To prevent particles from settling down, the mixture was mixed using a digital vortex mixer at 960 rpm. The mixture was subsequently coagulated with the addition of 2.06wt% calcium chloride at the volume ratio of 9:1. The clot was then water-bathed at 37°C for approximately 1 hour and was stored in the fridge at 5°C.

**Milli-spinner for drug delivery**

Water-soluble material was utilized to seal the slits of the milli-spinner. Food dye mimicking drug components was loaded into the milli-spinner's lumen through its front hole. When the lumen was filled with dye, the front hole of the milli-spinner was then sealed by the water-soluble material and the dye-encapsulated milli-spinner was put in the oven at 80°C for 1 hour for drying of the seal.

**Viscous fluid preparation**

Viscous fluids with a viscosity of 3.1 mPa·s, 5.4 mPa·s, and 7.0 mPa·s were created by mixing glycerol (Vegetable Glycerin, Raw Plus Rare Inc., USA) with water at the ratio of 7:3, 6:4 and 5:5. The mixture is stirred until well mixed. The viscosity of the fluid is measured using MCR 92 rheometer (Anton Paar, USA).

**Clot mechanical property characterization**

The blood clots with varied RBC contents were characterized using a universal test machine Instron 3344 (Instron, Inc., USA). The specimens were prepared by having blood mixtures filling 3.5 mm diameter fluorinated ethylene propylene (FEP) tubes and coagulated. For the whole blood and white fibrin clot tensile test, specimens were prepared by cutting clot chunks into clot strips. A strain rate of $0.01$ s$^{-1}$ was adopted for all tensile tests. The Young's modulus was fitted with the data to a strain of 0.1. The toughness of clots was calculated by taking the area under clots stress-strain curve using Origin 2024 (OriginLab Corporation).

**Milli-spinner suction demonstration by micro-PIV**

Micro-PIV was a method to measure the microscale flow profile of a plane. The measurement setup included a Nikon E400 Fluorescence Microscope (Nikon Corp., Japan), a high-power LED blue light source SOLIS-470C (Thorlabs, Inc., USA), a Phantom Miro-4M high-speed camera (Vision Research, Inc., USA), as well as the milli-spinner inside a water tank (36 mm × 10 mm × 10 mm). A 0.6× magnification changer attached to the high-speed camera, a 2× objective, and a Nikon blue excitation fluorescence filter cube DM505 B-2A were used. Fluorescent particles Dragon Green (1% solid, average size 15 μm, Bangs Laboratories, Inc., USA) were mixed into deionized water in a volume ratio of 1: 37.5. An exposure time of 600 μs and a frame rate of 1,000 fps were used, recording the fluorescent particle motions from the start of milli-spinner spinning until the stable state was reached. The raw data was processed based on an open source software PIVlab[51].

**Computational fluid dynamics (CFD) simulations**

In this study, CFD simulations were performed using COMSOL Multiphysics 6.1 (COMSOL Inc., USA) to qualitatively evaluate milli-spinner designs' effectiveness in generating suction and shear forces. The simulation utilized the turbulent k-epsilon model and modeled the spinning motion through the frozen rotor method. Mesh sensitivity analyses were conducted to ensure grid-independent solutions. It characterized the flow field around the milli-spinner induced by its spinning motion, including velocity, pressure, and shear force at different locations. The simulation was also implemented to cross-validate the Micro-PIV results of the milli-spinner design. The boundary effect of pipe size was investigated for each design. A parametric study was conducted to study the effect of fin length at different spinning frequencies.

**Visualization of whole blood clot and debulked clot microstructure through SEM**

The clots before and after debulking were prepared for visualization by SEM. The clots were fixed with 4% Paraformaldehyde, and 2% Glutaraldehyde in 0.1 M Sodium Cacodylate for 1 hour and placed in a 5°C fridge overnight. On the next day, the fixed clots were washed three times with 0.1M Sodium Cacodylate PH 7.4. buffer solution. Subsequently, the clots were post-fixed with 1% osmium tetroxide diluted in Sodium Cacodylate PH 7.4 for 1 hour at room temperature. Consecutively, the clots were dehydrated in ethanol (50%, 70%, 95%, and 2 times at 100%) for ten minutes, and the critical point dried. The clots were then coated with a layer of gold. The SEM images were taken using Zeiss Sigma SEM (Zeiss Inc., Germany) with GEMINI electron optical column.

**Animal care**

All animal procedures were approved by the institutional review board at Stanford. Eight swine (55-70 kg in size) were used in non-survival studies of the milli-spinner device. Swine were placed under general anesthesia by veterinarians trained in large animal care. Arterial access was obtained with an 8-French femoral sheath in the femoral and common carotid arteries, and the swine were heparinized with 5,000 units of heparin after sheath placement. All sheaths and catheters used for endovascular procedures were connected to a continuous infusion of pressurized heparinized saline throughout the experiment. Endovascular experiments were performed using conventional techniques with target arteries selected using continuous fluoroscopic guidance. At the conclusion of the procedure, the swine were euthanized by the veterinarians. Treated arteries and organs were harvested and placed in formalin for histologic analysis after the animals had expired.

**Histology procedure**

The kidney that the milli-spinner operated in was dissected and immediately fixed in 10% phosphate-buffered formalin for more than 24 hours. The operated vessels were dissected, trimmed, embedded in paraffin, cut into 5 µm sections, and stained with hematoxylin and eosin stain. The stained slides were checked using an Olympus Bx40 microscope. Histologic analysis was performed by a board-certified pathologist.

**Acknowledgements**

We would like to acknowledge Yamil Sanez and Robert Bennett for their assistance in the animal experiments. We would like to acknowledge Professor Juan G. Santiago for support on Micro-PIV setup and the useful insights on Micro-PIV and CFD results. We would like to acknowledge Six Oliva Skov for helping construct the Micro-PIV setup. We would also like to acknowledge Dr. Lu Lu for his contribution on CFD simulations. The authors acknowledge the support of institutional grants by the Wu Tsai Neurosciences Institute, Stanford-Coulter Translational Research Grants, Stanford Medicine Catalyst, and NIH Shared Instrumentation grant (S10RR026714).

**Author contribution**

R.R.Z. designed and supervised the research; R.R.Z., Y.C., Q.L., S.W., B.P, J.J.H. performed the *in-vivo* experiment; Y.C., Q.L., S.W realized the *in-vitro* experiments and analyzed the data; Q.L. realized the CFD simulations; D.S. performed histology analysis. R.R.Z., Y.C., Q.L., S.W., J.J.H., and B.P. wrote the paper. All authors contributed to the discussion.

**Competing interest**

Two PCT applications have been filed on the reported technology.

**Data availability**

The authors declare that the data that support the findings of this study are available from the corresponding author upon the reasonable statement.